\documentclass[doublecol]{epl2}

\title{ Tonks-Girardeau gas, super-Tonks-Girardeau gas, and bound
states of one-dimensional bosons in a hard-wall trap}
\shorttitle{TG, STG and Bound State of 1D Bosons in a Hard Wall Trap} 

\author{Hua Li\inst{1} \and Taifeng Liu\inst{1} \and Yaojiang Hao\inst{2} \and Yunbo Zhang\inst{1}}
\shortauthor{Hua Li \etal }

\institute{
  \inst{1} Institute of Theoretical Physics and Department of Physics, Shanxi University, Taiyuan, Shanxi 030006, China\\
  \inst{2} Department of Physics, University of Science and Technology Beijing, Beijing 100083, China
}


\abstract{ We investigate the Bose gas with repulsive or attractive
interactions between atoms in the scheme of Bethe Ansatz equation in
a hard wall trap. Three typical quantum phases in the current
research of 1D interacting cold atoms are clarified in terms of the
energy spectrum, single particle density distribution and
two-particle correlation function. We identify two matching points
in the phase diagram, i.e. the TG and STG gas show the same profiles
at the strongly interacting point $-1/\gamma=0$, and in the weakly
interacting limit $\gamma=0$ the ground states TG and BS join to
each other smoothly.}

\begin{document}

\maketitle

Recently experiments and theories have made a great progress in one
dimensional quantum gases \cite{Cazalilla}. Thanks to the Feshbach resonance as well
as the confinement-induced resonance (CIR)
\cite{Olshanii,Bergeman,Sinha}, the interaction strength between
atoms can be tuned exactly. The first breakthrough in experiment is
the realization of Tonks-Girardeau (TG) gases
\cite{Kinoshita,Paredes} of bosons with strongly repulsive
interaction. The atoms in TG gas stay in ground state and the wave
function and energy was found by Girardeau in 1960 using a
Fermi-Bose (FB) mapping \cite{Girardeau}, leading to fermionization
of many properties of this Bose system. The many-body properties
were determined explicitly for a bosonic Tonks-Girardeau gas
confined in the potential of a harmonic trap \cite{Goold,Lu} or a
hard wall trap \cite{Yin} with a tunable $\delta$-function barrier
at the trap center and related work has been done for Bose-Ferimi
mixture \cite{Lelas,Lu} and finite interacting strength
\cite{Deuretzbacher,Zollner,Hao06}.

\begin{figure}
\includegraphics[width=3in]{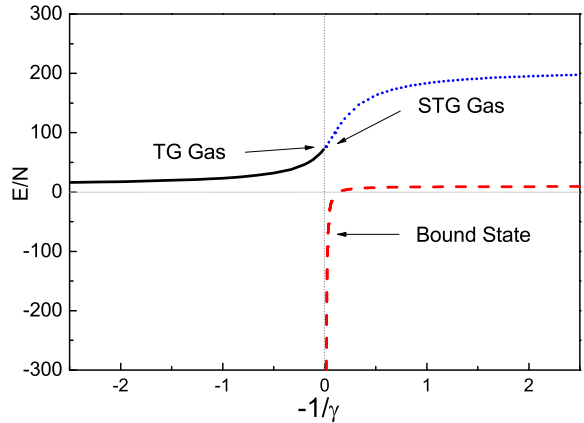}
\caption{(Color online) The energy spectrum of $N=4$ bosons for the
ground states and one of the excited states in the full interacting
regime. Quantum phases of TG, STG and Bound State are clearly
classified into three branches solution of Bethe Ansatz Equations.}
\label{fig.1}
\end{figure}

The second important experiment is the realization of the super
Tonks-Girardeau (STG) gas \cite{Haller}, where the Cesium atoms are
prepared in a long-lived, strongly interacting, excited state.
Unlike the strong repulsively interacting TG gas, the interaction of
bosons in STG state is attractive. Although the TG gas has proved
very successful both in theories and experiments, bosons with
attractive interaction start to attract more and more attention only
recently. In fact, if the interaction is finite and negative, the
ground state is McGuire's cluster state \cite{McGuire, Tempfi}, which is a
bound state and would decay quickly via molecular channels. In the
original proposal, the STG gas-like state corresponds to a highly
excited Bethe state in the integrable interacting Bose gas for which
the bosons acquire hard-core behavior
\cite{Batchelor,Astrakharchik}. Experimentally the STG gas was
realized through a sudden quench of the interaction parameter
\cite{Chen1}. On the other hand, the ground state of a strongly
attractive spin-1/2 Fermi gas can be effectively described by the
STG gas \cite{Chen2}.

Exact solutions are available only in very few limiting cases. 1D
Bose gas with arbitrary repulsive interaction strength was solved by
Lieb and Liniger \cite{Lieb} using the Bethe ansatz method in which
particles are constrained by periodic boundary conditions on a line
of length $L$. As far as the trapping of the atoms is concerned, a
solution was given for two $\delta$-interacting particles in a
harmonic trap in \cite{Busch} and the STG wave function for $N = 2$
particles is found to be identical to the hard-sphere bose gas
\cite{Girardeau2} given exactly in terms of a parabolic cylinder
function. It is very interesting to consider what happens to the
above mentioned three branch solutions of 1D Bose gas when the atoms
are confined in a hard wall trap, i.e., under the open boundary
condition. Interacting bosons in a hard-wall trap is one of the few integrable examples of
interacting many-body models and its exact solution has been obtained with the
superposition of Bethe's wave function for the repulsive interaction in a
seminal paper by \cite{Gaudin}. For a finite number of bosons and
finite system size, boundary effects are expected to be pronounced at low
temperature. Indeed, significantly different quantum effects should be exhibited
by a finite number of bosons confined in a finite hard wall box.

Here we study the property of a few bosons trapped in 1D
hard wall trap by numerically solving the Bethe Ansatz equations in
the full interacting regime. We compare the energy of the system,
the one body density matrix and the two-particle correlation
function both in the ground state for both repulsive and attractive
interactions and in the excited state for attractive interaction.
The transition probability from the TG gas to the STG gas is
examined through a sudden switch of interaction. 
Typical features are shown to be quite
different for the three branches of Bethe Ansatz solution, which may serve as a
guide line for the experiments with hard wall boundary condition.

Consider $N$ bosonic particles of mass $m$ in the 1D hard wall of
width $L$ with $\delta $ function interaction. The stationary
behavior of the many-body wave function
$\Psi (x_1, x_2, \cdots  x_N))$ is described by the Schr\"odinger equation%
\begin{equation}
-\frac{\hbar^2}{2m}\sum\limits_{i=1}^{N}\frac{{\partial
^{2}}}{{\partial x_{i}^{2}}}\Psi +g_{1D}
\sum\limits_{i,j(i<j)}^{N}\delta \left( x_{i}-x_{j}\right) \Psi
=E\Psi . \label{Ha}
\end{equation}%
The interaction strength $g_{1D}=-2 \hbar^2 / m a_{1D}$ and $a_{1D}=
a_\perp \left( C-a_\perp/a_{3D} \right)$ is the 1D scattering
length. The tightly confining wave-guides in experiments are related
to the theoretical 1D models through the so-called Confinement
Induced Resonance (CIR). Across the resonance point $a_\perp
/a_{3D}=C=1.0326$, the $g_{1D}$ can be changed from $-\infty $ to
$+\infty$ by (i) varying the transverse width of the waveguide
$a_\perp=\sqrt{\hbar/m \omega_\perp}$ where $\omega_\perp$ is the
frequency of the transverse confinement, and/or (ii) tuning the 3D
scattering length $a_{3D}$ by means of a magnetically induced
Feshbach resonance. In the paper we use the dimensionless
Tonks-Girardeau parameter $\gamma =c/n$ with $c=m g_{1D}/\hbar^2$ is
the Lieb-Lininger interaction parameter and $n=N/L$ is the particle
density, to characterize the interaction strength between the
bosons. For convenience we set $\hbar =2m=1$ in the following.

The Lieb-Lininger model (\ref{Ha}) is exactly solvable for both
repulsive ($c>0$) and attractive ($c<0$) interaction. The basic idea
is that we first try to find the wave fuction in the local region
$R_{1}: 0\leq x_{1}\leq x_{2}\leq \cdots \leq x_{N}\leq L $, i.e.
assume it takes the form of a superposition of plane waves with wave
vectors $k_{P_j}$, $\varphi \left( x_{1},x_{2},\cdots ,x_{N}\right)
=\sum\nolimits_{P,r_{1},\cdots ,r_{N}}A_{P}\exp \left(
i\sum\nolimits_{j}r_{j}k_{P_{j}}x_{j}\right) $, where $r_{j}=\pm 1$
represent the particle moving to the right or to the left and $A_P$
are coefficients to be fixed by the boundary conditions.
$P_{1},P_{2},\cdots P_{N}$ is the permutation of the $1,2,\cdots N$,
$\sum\nolimits_{P}$ is the summation over all the permutations.
Under the open boundary condition, $\varphi \left( 0, x_{2},\cdots
,x_{N}\right) =\varphi \left( x_{1}, x_{2},\cdots ,L\right) =0$, the
Bethe Ansatz equations are
\begin{equation}
\exp \left( i2k_{j}L\right) =\prod\limits_{l=1\left( \neq j\right) }^{N}%
\frac{i\left( k_{l}+k_{j}\right) -c}{i\left( k_{l}+k_{j}\right) +c}
\frac{i\left( k_{l}-k_{j}\right) +c}{i\left( k_{l}-k_{j}\right) -c}
\label{Beth}
\end{equation}%
with $j=1,2, \cdots N$. In the region of space under consideration
$R:  0\leq x_{1},x_{2},\cdots ,x_{N}\leq L$, the wave function is
represented as a summation over all permutations of coordinates
\begin{eqnarray}
\Psi \left( x_{1},\cdots ,x_{N}\right) &=&\sum\limits_{P}\varphi
\left(
x_{P_{1}},x_{P_{2}},\cdots ,x_{P_{N}}\right)  \nonumber \\
&&\times  \theta \left( x_{P_{1}}<x_{P_{2}}<\cdots <x_{P_{N}}\right)
. \label{WF}
\end{eqnarray}%
with
\begin{equation}
\theta \left( x_{P_{1}}<\cdots <x_{P_{N}}\right) =\theta \left(
x_{P_{N}}-x_{P_{N-1}}\right) \cdots \theta \left(
x_{P_{2}}-x_{P_{1}}\right)
\end{equation}%
where the step function $\theta \left( x-y\right) $ is 0 for $x<y$,
while it is 1 for $x>y$.

The quasimomenta $\left\{ k_{j}\right\} $ can be solved from the Bethe Ansatz equations (\ref%
{Beth}). The energy and the total momentum of the system are thus
$E=\sum_{j=1}^{N}k_{j}^{2}$ and $k=\sum_{j=1}^{N}k_{j}$,
respectively. Correspondingly the wave function $\Psi$ is known as
(\ref{WF}), on which all physical quantities can be calculated.
Taking the logarithm of (\ref{Beth})
\begin{eqnarray}
k_{j}L &=&I_{j}\pi -\sum_{l=1\left( \neq j\right) }^{N}\arctan
\frac{\left(
k_{j}-k_{l}\right) }{c}  \nonumber \\
&&-\sum_{l=1\left( \neq j\right) }^{N}\arctan \frac{\left(
k_{j}+k_{l}\right) }{c}, \label{log}
\end{eqnarray}
leads us to $N-1$ coupled transcendental equations that determine
both the exact $N$-particle ground state and excited states of the
problem. It is clear that the solution of $\left\{ k_{j}\right\} $
depends critically on the choices of the quantum numbers $\left\{
I_{j}\right\} $, which is a set of integers, and the initial values
of $\left\{ k_{j}\right\} $. Here we focus on characteristic
properties of three typical cases in the current research of 1D
interacting cold atoms, all of which are solutions of equation
(\ref{log}) with $I_{j}=j$. They are: (i) \textbf{TG Branch:} For
$c>0$, the ground state solution of $\left\{ k_{j}\right\} $  is a
set of real numbers. This solution approaches to the TG gas when
$c\rightarrow +\infty $ with $k_j=j \pi/L$. (ii) \textbf{Bound State
Branch:} For $c<0$, the $\left\{ k_{j}\right\} $ solution for the
ground state is complex which is known as the string solution just
like the dimer and trimer states of bosons with periodic boundary
condition \cite{Sakmann,Muga}. For example, the quasimomenta
$\left\{ k_{j}\right\} $ in the case of $N=4$ are respectively
$k_{1,2}=\alpha _{1}\ \pm i\Lambda _{1}$, $k_{3,4}=\alpha _{2}\pm
i\Lambda _{2}$, which form pairs with conjugated partners. In the
limit of strong attractive interaction, the $N$-string solution
takes the form $k_{j}=\pi /NL+i(N+1-2j)(c/2+\delta _{j})/L$ where
$\delta _{j}$ is a set of small numbers which exponentially decay to
zero when $c\rightarrow -\infty $ and the energy of ground state is
$E/N=k^{2}/N^2L^2-c^2(N^{2}-1)/12L^2$. (iii) \textbf{STG Branch:}
The real solutions $\left\{ k_{j}\right\} $ of equation (\ref{log})
for $c<0$ describe one of the highly excited states of the system.
In the strongly attractive interaction limit $c\rightarrow -\infty
$, it corresponds to the STG gas.

\begin{figure}
\includegraphics[width=3in]{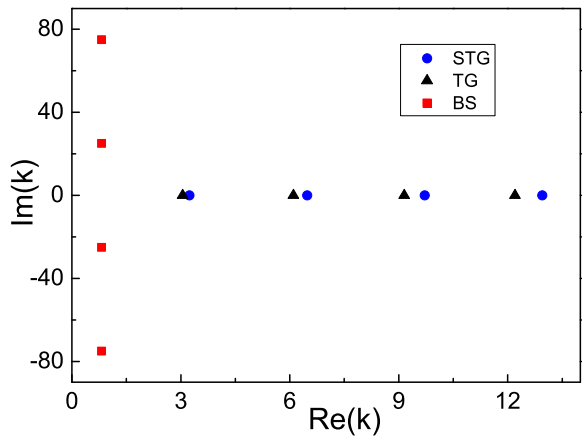}
\caption{(Color online) The quasi momentum spectrum of $N=4$ bosons
of TG, STG and Bound State branches for interaction constant
$\left\vert \protect\gamma \right\vert =50$.} \label{fig.2}
\end{figure}

\begin{figure}
\includegraphics[width=3in]{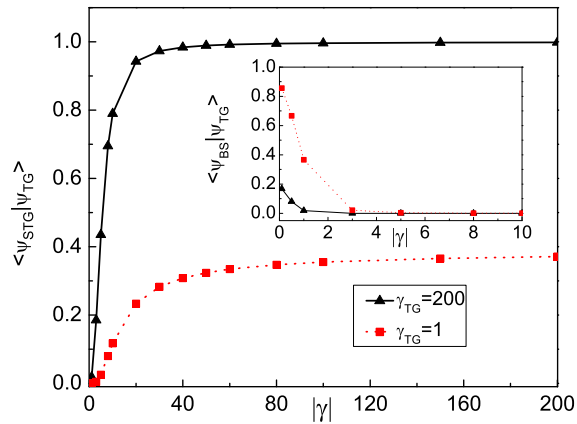}
\caption{(Color online) The transition probability from TG branch to
STG branch for $N=4$ bosons. The system is initially prepared in TG
gas with $\gamma=200$ and Lieb-Lininger gas with $\gamma=1$. Inset
shows the corresponding transition probability to the BS branch.
\label{fig.3}}
\end{figure}

We show the energy spectrum of $N=4$ bosons for the three branches
in the hard wall trap of width $L=1$ in Fig. 1. As in the case of
homogeneous problem of three particles with periodic boundary
conditions \cite{Muga} and two particles in a harmonic trap
\cite{Busch,Muth}, the spectrum is quite different depending on the
sign of the interaction parameter $\gamma$ or $c$ (we use
$-1/\gamma$ in all figures in this paper). However, two matching
points can be identified for the three solutions. In the
non-interacting limit the ground states for the weakly repulsive (TG
Branch) and weakly attractive (Bound State Branch) interacting
systems join smoothly at the left and right end of the axis
($\gamma=0$) with $E/N=\pi^2/L^2$. On the other hand, in the
strongly interacting limit the TG Branch adiabatically connects to
the STG Branch at the origin of the axis ($-1/\gamma=0$) with energy
$E/N=\pi^2 (N+1)(2N+1) /6L^2$.

The solutions of quasi momentum $\left\{ k_{j}\right\} $ for the
three cases are presented in the complex plane of $k$ in Figure 2
for a large interacting parameter $\left\vert \gamma \right \vert
=50$. The values are real for TG and STG branches, which become
closer and closer when approaching the strong interacting point
$-1/\gamma=0$. Unlike the case of periodic boundary condition
\cite{Chen1}, they are not any more symmetric about $k=0$. By
assigning the initial value of $\left\{ k_{j}\right\} $ a complex
number, one obtains the string solutions for BS branch with a rather
small real part and the imaginary parts lie symmetrically along the
axis Im(k), indicating the conjugacy.

\begin{figure*}[tbp]
\includegraphics[width=7in]{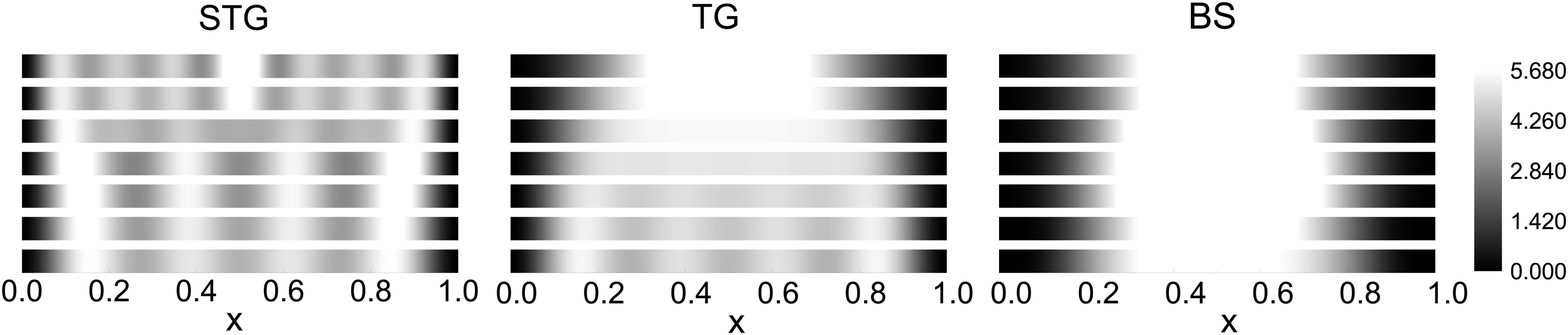}
\caption{(Color online) The single particle density profiles for the
three typical quantum phases of 1D interacting cold atoms. Left,
middle and right panels show the gray scale plots for STG, TG and BS
branches respectively, for seven gradually increasing interaction
parameters and N = 4.
Parameters from the top down are $\gamma =-0.1,-1,-2.5,-5,-12.5,-25,-100$ for STG, $\gamma %
=0.1,1,2.5,5,12.5,25,100$ for TG, and $\gamma
=-0.1,-0.5,-1,-2,-3,-4,-5$ for the BS branches.}
\label{fig4}
\end{figure*}

As shown in Figure 1, the energy changes smoothly from TG gas to STG
gas when tuning the interaction parameter $-1/\gamma$ across the
strongly interacting point from $-0$ to $+0$. This actually insures
the deterministic preparation of the STG gas from the TG gas after a
sudden change of interaction parameter via CIR. However, for the
general case of an interacting bosonic system, this quench dynamics
is complicated because the resulting state is a linear combination
of eigenstates of the new Hamiltonian with $-1/\gamma>0$. The
transition probability to other states, such as the bound states, is
non-negligible for weakly interacting final state. Figure 3 shows
the quenching dynamics of this interacting boson model, where the
initial state is prepared in the TG Branch with $\gamma =200$ or
$\gamma =1$, respectively. The transition probability to the final
state in STG branch, characterized as $\left\vert \left\langle \Psi
_{STG}\mid \Psi_{TG}\right\rangle \right\vert ^{2}$, is plotted for
varying parameter $\left \vert \gamma \right \vert$. The result
confirms the metastable feature of the STG gas against collapsing
into the bound state in the strongly attractive interaction regime.
On the other hand, one has less probability to achieve the STG gas
if the initial state is weakly repulsive. Inset in Figure 3 shows
the transition probability to the bound state branch $\left\vert
\left\langle \Psi _{BS}\mid \Psi_{TG} \right\rangle\right\vert
^{2}$, for the same initial states. It is more likely that one will
arrive at the bound state of small $\left \vert\gamma\right \vert$
if  the system is initially prepared in the weakly interacting
Lieb-Lininger gas.

Since single-particle density profiles normalized to $N$ depends
crucially on the many-body wave function $\Psi$, one might
anticipate to more easily recognize
the distinction between the three quantum phases from their density distribution defined as%
\begin{equation}
\rho \left( x\right) =N\int_{0}^{L}dx_{2}\cdots
\int_{0}^{L}dx_{N}\left\vert \Psi \left( x,x_2\cdots ,x_{N}\right)
\right\vert ^{2}. \label{rho}
\end{equation}%
Physically, $\rho \left( x\right) $ represents the probability that
the bosons occupy the position $x$, which is accessible
experimentally by means of the time of flight imaging technique. The
single-particle density profiles for the three branches solution are
illustrated in gray scale plot in Figure 4 for several gradually
increasing interaction parameters and $N=4$. The left panel shows
that in the excited state of STG branch the density of weakly
interacting bosons exhibits seven peaks due to the slightly
attraction between atoms, which develops gradually into four in the
bottom plot when approaching the strong interacting STG regime.
 The density in the ground state of repulsive interacting bosons evolves in a
quite different way as shown in the middle panel, however, merges
into an identical four-peak structure in the strongly interacting
point. This feature indicates the fermionization of bosonic atoms
for either repulsive or attractive interaction for large value of
$\left \vert \gamma \right \vert$. On the right panel, atoms in the
ground bound state display always in a single peak, even for very
large interaction parameter. We observe again the two matching
points at $\gamma=0$ and $-1/\gamma=0$, as can be easily seen from
the almost identical profiles at the bottom two plots of STG and TG
branches and those at the top two plots of TG and BS branches.

\begin{figure}[tbp]
\includegraphics[width=3in]{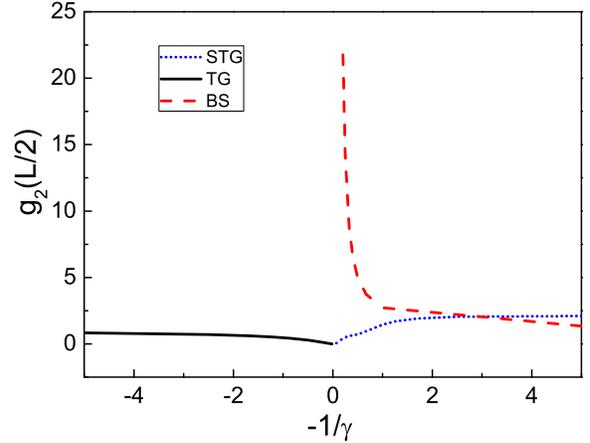}
\caption{(Color online) The two-particle correlation function of TG,
STG and Bound State branches in unit of $g_{2}\left(
L/2,\protect\gamma =0\right)$.} \label{fig5}
\end{figure}

Another interesting feature of 1D exactly solvable model is the
two-particle correlation function defined as
\begin{eqnarray}
g_{2}\left( x\right) &=&N\left( N-1\right) \int_{0}^{L}dx_{3}\cdots
\int_{0}^{L}dx_{N}\nonumber  \\
&&\times\left\vert \Psi \left( x,x,x_3\cdots ,x_{N}\right)
\right\vert ^{2}, \label{g2}
\end{eqnarray}
which may be used to identify and classify the quantum phases of 1D
Bose gas. It is of particular importance for the studies of
coherence properties of atom lasers produced in 1D waveguide.
$g_{2}\left( x\right)$ denotes the probability that two successive
measurements will find an atom at the same position $x$. The factor
of $N(N-1)$ in Eq. (\ref{g2}) reflects there are $N(N-1)$ ways of
choosing the $N-2$ integral variables out of $N$, just as we have
$N$ ways of choosing the $N-1$ integral variables out of $N$ in the
density distribution Eq. (\ref{rho}). In the Fig. 5, we show the two
particle correlation function for 4 bosons in the position $x=L/2$
by numerically solving the Bethe Ansatz equations and doing the
multiple integral for wave functions. We emphasize that the method
used here is different from the approach in \cite{Chen1}, where the
correlation function has been obtained by numerically solving the
integral equation in the thermodynamic limit. It can be perceived
that generally the STG branch is more strongly correlated than the
TG branch, especially $g_2(L/2)=2$ for very small value of
$\left\vert \gamma \right\vert $. At the strong interacting point,
the correlation vanishes for both STG and TG branches, $g_2(L/2)=0$,
in agreement with the result of periodic boundary condition in
thermodynamic limit \cite{Chen1}. On the contrary, the correlation
function for the ground bound state with strong attractive
interaction is significantly larger than the STG and TG gas, while
it decreases rapidly to the result of non-interacting system,
$g_2(L/2)=1$.

In summary, we have theoretically studied the ground state and
excited state of a 1D Bose gas with arbitrary interaction strength
in a hard wall trap. We show in the energy phase diagram the three
quantum phases TG, STG and BS are closely connected to each other in
the weakly and strongly interacting limits. We also calculate the
density distribution and correlation function in the above cases and
conclude that the STG gas is a strongly correlated state.

\acknowledgments This work is supported by the NSF of China under
Grant No. 11074153, the National Basic Research Program of China
(973 Program) under Grant No. 2010CB923103, the NSF
of Shanxi Province, Shanxi Scholarship Council of China, and the
Program for New Century Excellent Talents in University (NCET).\\

\end{document}